\documentclass{ifacconf}

\usepackage{graphicx}      
\usepackage{natbib}        
\usepackage{epstopdf}
\usepackage{amsmath}
\usepackage{amsfonts}
\usepackage{dsfont}
\usepackage{amssymb}
\usepackage{color}

\newtheorem{defin}{{Definition}} 
\newtheorem{remark}{{Remark}} 
\newtheorem{theorem}{Theorem}

\newtheorem{proposition}{Proposition}
\begin{document}
\begin{frontmatter}

\title{Tracking Control of Fully-actuated Mechanical port-Hamiltonian Systems using Sliding Manifolds  and Contraction}



\thanks[footnoteinfo]{The first author thanks to the Mexican Consul of Science and Technology and the Government of the State of Puebla for the grant assigned to CVU  $386575$.}

\author[First,Second]{Rodolfo Reyes-B\'aez}
\author[First,Second]{Arjan van der Schaft} 
\author[First,Third]{Bayu Jayawardhana}

\address[First]{Jan C. Willems Center for Systems and Control,\\ University of Groningen, The Netherlands}

\address[Second]{Johann Bernoulli Institute for Mathematics and Computer Science, University of Groningen, P.O. Box 407, 9700 AK, Groningen, The Netherlands (\{r.reyes-baez, a.j.van.der.schaft\}@rug.nl)}

\address[Third]{Engineering and Technology Institute Groningen (ENTEG), University of Groningen,  Nijenborgh 4, 9747AG, The Netherlands (b.jayawardhana@rug.nl)}

\begin{abstract}                
In this paper, we propose a novel trajectory tracking controller for fully-actuated mechanical port-Hamiltonian (pH) systems, which is based on recent advances in contraction-based control theory. Our proposed controller renders a desired sliding manifold (where the reference trajectory lies) attractive by making the corresponding error system partially contracting. Finally, we present numerical simulation results where a SCARA robot is commanded by our proposed tracking control law. 
\end{abstract}

\begin{keyword}
Trajectory tracking control, port-Hamiltonian systems, sliding manifold, differential Lyapunov theory, partial contraction
\end{keyword}

\end{frontmatter}

\section{Introduction}

The control of electro-mechanical (EM) systems is a well-studied problem in control theory literature. Using Euler-Lagrange (EL) formalism for describing the dynamics of EM systems, many control design tools have been proposed and studied  to solve the stabilization/set-point regulation problem. Recent works include the passivity-based control methods which are expounded in \cite{ortega2013passivity} and references therein. However, for motion control/output regulation problem (which includes trajectory tracking and path-following problems), the use of EL formalism in the control design is relatively recent and the problem is solved based on passivity/dissipativity theory for nonlinear systems (\cite{willem1972}). We refer interested reader on the early work of tracking control for EL systems in \cite{slotineli} and recent works in \cite{kelly2006control,jayawardhana2008}.   



As an alternative to the EL formalism for describing EM systems, port-Hamiltonian (pH) framework has been proposed and studied (see also the pioneering work in \cite{vanderschaft1995} and the recent exposition in \cite{pHbook}), which has a nice (Dirac) structure, provides port-based modeling and has physical energy interpretation. For the latter part, the energy function can directly be used to show the dissipativity and stability property of the systems. The port-based modeling of pH systems is modular in the sense that we can interconnect pH systems through their external ports with a power-preserving interconnection that still preserve the pH structure of the interconnected system. 

Using the pH framework, a number of control design tools have been proposed and implemented for the past two decades. For solving the stabilization and set-point regulation problem of pH systems, we can apply, for instance, the standard proportional-integral (PI) control (\cite{jayawardhana2007}), Interconnection and Damping Assignment Passivity-Based Control (IDA-PBC) approach (\cite{escobar}) or Control by Interconnection (CbI) method (\cite{castanos}) (among many others). 

However, for motion control problem, where the reference signal can be time-varying, it is not straightforward to design control laws for such pH systems that still provides an insightful energy interpretation of the closed-loop system. For example, it is not trivial to obtain an incremental passive system\footnote{This concept generalizes the usual notion of passivity and is suitable for output regulation problem of non-constant signals, see \cite{jayawardhana2006}, \cite{pavlov2006}} via a controller interconnected with the pH system. One major difficulty is that the external reference signals can induce both the closed-loop system and total energy function to be time-varying. In this case, the closed-loop system may not be dissipative, or if it is a time-varying dissipative system, the usual La-Salle invariance principle argument can no longer be invoked for analyzing the asymptotic behavior.  


In order to overcome the loss of passivity in the trajectory tracking control of pH systems, a pH structure preserving error system was introduced in \cite{fujimoto} which is based on generalized canonical transformations. In \cite{fujimoto}, necessary and sufficient conditions for passivity preserving are given. Once in the new canonical coordinates, the pH error system can be stabilized with standard passivity-based control methods. 

In \cite{dirksz2010adaptive}, the previous approach is extended to an adaptive control one. 
In \cite{romero2015passivity} and similarly in \cite{lupe}, a generalized canonical transformation is used to obtain a particular pH system which is \emph{partially linear} in the momentum with constant inertia matrix. The control scheme is then proposed to give a pH structure for the closed-loop error system. Although solving partial differential equations that correspond to the existence of such transformation is not trivial, some characterizations of this canonical transformation is presented in \cite{venkatraman2010speed} for a specific class of systems. In \cite{iran}, the so-called \emph{timed-IDA-PBC} was introduced, where the standard IDA-PBC method for stabilization is adapted in such a way that it can incorporate tracking problem through a modification in the IDA-PBC matching equations; albeit it may easily lead to a non-tractable problem in solving a set of complex PDE. 
Finally, in a recent paper \cite{tracking2016}, a trajectory tracking control for standard pH systems without dissipation is proposed, using a similar change of coordinates as in \cite{slotineli} where the  the \emph{Coriolis} term is defined explicitly in the Hamiltonian domain. This approach is motivated by the work of \cite{Arimoto}.


As an alternative to the use of passivity-based control method for solving motion control problems, we propose in this paper a contraction-based control method for fully-actuated pH systems. The main idea is to combine the recent result in tranverse exponential stability (\cite{andrieu2016}), where an invariant manifold is made attractive through a contraction-based control law, and the sliding-manifold approach (\cite{spong}), where we ensure that on the invariant manifold, the trajectory converges to the desired reference trajectory via an additional control law. 
There already exist different approaches to get attractivity of the sliding manifold $\mathcal{S}$, such as, sliding mode control (\cite{utkin}), singular perturbations techniques (\cite{spong}),  passivity (\cite{slotineli}),  Immersion \& Invariance (I\&I) approach (\cite{forniII}) and reduction methods (\citealp{maggiore}). 

Generally speaking, we first construct an error system using backstepping method such that the closed-loop error system (in the sense of \cite{fujimoto}) has a time-varying pH structure as in (\cite{lupe}) but not necessarily with a constant inertia matrix. In order to show the convergence of the system's trajectory to the time-varying reference trajectory, the state space is extended with the incorporation of a \emph{virtual system} where the latter system admits both the system's trajectory, as well as, the reference trajectory as its solution. By using the transverse exponential stability results as in (\cite{andrieu2016}) or the \emph{partial contraction} as in (\cite{wang}), the contraction of the virtual system in the extended system implies that state trajectory converges exponentially to the desired one.  




\section{Preliminaries}\label{sII}
\subsection{Control design using sliding manifolds}
Let $\mathcal{X}$ be the state-space with tangent bundle  $T\mathcal{X}$  of a nonlinear system given by
\begin{equation}
\dot{\vec{x}}=\vec{f}(\vec{x},t)+\vec{G}(\vec{x},t)\vec{u}
\label{eq:controlsystemSliding}
\end{equation}
where $\vec{G}(\vec{x},t)=[\vec{g}_1(\vec{x},t),\cdots,\vec{g}_n(\vec{x},t)]$  has  full rank and  $\vec{x}\in\mathcal{X}\subset\mathds{R}^{N}$ the solution to \eqref{eq:controlsystemSliding},  $\vec{f}, \vec{g}_k$ are smooth vector fields on $\mathcal{X}\times\mathds{R}_{\ge 0}$ for $k\in\{1,\dots,n\}$  and $\vec{u}=[u_1,\cdots,u_n]^{\top}$ the control input. 
Recall the definitions of invariant and sliding manifolds.
\begin{defin}[\cite{spong}]
	 The set ${\mathcal{S}}\subset\mathcal{X}$ is said to be an \emph{invariant manifold} for system \eqref{eq:controlsystemSliding} if whenever $\vec{x}(t_0)\in{\mathcal{S}}$, implies that $\vec{x}(t)\in\mathcal{S}$, for all $t>t_0$.	 
\end{defin}
\begin{defin}[\cite{sira1988differential}]
	A \emph{sliding  manifold} for system \eqref{eq:controlsystemSliding} is a subset of the state space, which is the intersection of $n$ smooth $(N-1)$-dimensional manifolds,
	\begin{equation}
	{\Omega}(t)=\{\vec{x}\in\mathcal{X}: \boldsymbol{\sigma}(\vec{x},t)=0\}
	\label{eq:slidingmanifold}
	\end{equation}
	where $\boldsymbol{\sigma}(\vec{x},t)=[\sigma_1(\vec{x},t),\dots, \sigma_n(\vec{x},t) ]^{\top}$ is the \emph{sliding variable} with $\sigma_i$ a smooth function $\sigma_i:\mathcal{X}\times\mathds{R}_{\ge 0}\rightarrow \mathds{R}$. 	
\end{defin}
It is assumed that ${\Omega}(t)$  is locally an $(N-n)$-dimensional, sub-manifold of $\mathcal{X}$.
The smooth control vector $\vec{u}_{eq}$, known as the \emph{equivalent control}, renders the manifold $\Omega$ to an invariant manifold $\mathcal{S}$ of \eqref{eq:controlsystemSliding} (\cite{utkin}). If $\text{rank}\{L_{\vec{G}}\boldsymbol{\sigma}\}=n$, the equivalent control is the well defined solution to the following invariance conditions
\begin{equation}
\boldsymbol{\sigma}(\vec{x},t)=0, \quad \dot{\boldsymbol{\sigma}}(\vec{x},t)=0, 
\label{eq:invariancecontidionsSliting}
\end{equation}
uniformly in t. The dynamical system $\dot{\vec{x}}=\vec{f}(\vec{x},t)+\vec{G}(\vec{x},t)\vec{u}_{eq}(\vec{x},t)$ is said to describe the \emph{ideal sliding motion}.

Using sliding manifolds in control design has as goal, designing a suitable control scheme  $\vec{u}=\vec{u}_{eq}+\vec{u}_{at}$, such that $\vec{u}_{eq}$ renders $\Omega(t)$ to an invariant sliding manifold, under invariance conditions \eqref{eq:invariancecontidionsSliting}, and $\vec{u}_{at}$ makes to the invariant manifold attractive.

\subsection{Contraction analysis and differential Lyapunov theory}
System \eqref{eq:controlsystemSliding} in closed-loop with $\vec{u}$, denoted by
\begin{equation}
\dot{\vec{x}}=\vec{F}(\vec{x},t),
\label{eq:systemcontraction1}
\end{equation}
is be called \emph{contracting}, if initial any pair of solution $\vec{x}_1$ and $\vec{x}_2$ converges to each other, with respect to a \emph{distance}. In this paper, for contraction analysis, we adopt the approach given in \cite{forni}.

The \emph{prolonged system} (\cite{crouch}) of \eqref{eq:systemcontraction1} corresponds to the original system \emph{together} with its \emph{variational system}, that is the system 
\begin{equation}
\begin{split}
\left\{ \begin{array}{lcc}
\dot{\vec{x}}=\vec{F}(\vec{x},t) \\
\dot{\delta\vec{x}}=\frac{\partial \vec{F}}{\partial\vec{x}}(\vec{x},t)\delta\vec{x}\\
\end{array}
\right.
\end{split}
\label{eq:prolongedsystemcontraction}
\end{equation}
with $(\vec{x},\delta\vec{x},t)\in T\mathcal{X}\times\mathds{R}_{\geq 0}$. A \emph{differential Lyapunov function} $V:T\mathcal{X}\times\mathds{R}_{\ge 0}\rightarrow \mathds{R}_{\geq 0}$ satisfies the bounds
\begin{equation}
c_1\| \delta\vec{x}\|_{\vec{x}}^{p} \leq V(\vec{x},\delta\vec{x},t) \leq c_2\| \delta\vec{x}\|_{\vec{x}}^{p},
\label{eq:finslerlyapunov}
\end{equation}
where $c_1,c_2\in\mathds{R}_{>0}$, $p$ is some positive integer and $\|\cdot \|_{\vec{x}}$ is a Finsler structure. The role of \eqref{eq:finslerlyapunov} is to measure the distance of any tangent vector $\delta\vec{x}$ from $\vec{0}$. Thus, \eqref{eq:finslerlyapunov} can be understood as a classical Lyapunov function for the linearized dynamics with respect to the origin in $T \mathcal{X}$.

\begin{theorem}\label{eq:lyapunovcontraction}
	Consider the prolonged system \eqref{eq:prolongedsystemcontraction}, a connected and forward invariant set $\mathcal{D}$, and a strictly increasing function $\alpha:\mathds{R}_{\geq 0}\rightarrow \mathds{R}_{\geq 0}$. Let $V$ be a differential Lyapunov function satisfying
	\begin{equation}
	\dot{V}(\vec{x},\delta \vec{x},t)	\leq -\alpha(V(\vec{x},\delta \vec{x},t)) 
	\label{eq:finslerlyapunovinequality}
	\end{equation}
	for each $(\vec{x},\delta\vec{x})\in T\mathcal{X}$ and uniformly in $t\in\mathds{R}_{\geq 0}$. Then, \eqref{eq:systemcontraction1} contracts $V$ in $\mathcal{D}$. $V$ is called the \emph{contraction measure}, and $\mathcal{D}$ \emph{the contraction region}. 
\end{theorem}
\begin{remark}\label{remark:incremental}
	Contraction of \eqref{eq:systemcontraction1} is guaranteed by \eqref{eq:finslerlyapunov} and \eqref{eq:finslerlyapunovinequality}, with respect to the distance induced by the Finsler measure $\|\cdot \|_{\vec{x}}$, through integration. As direct consequence (\cite{forni}), system \eqref{eq:systemcontraction1} is incrementally
	\begin{itemize}
		\item   \emph{stable} on $\mathcal{D}$ if $\alpha(s)=0$ for each $s\geq 0$;
		\item  \emph{asymptotically stable} on $\mathcal{D}$ if $\alpha$ is a strictly increasing;
		\item  \emph{exponentially stable} on $\mathcal{D}$ if $\alpha(s)=\beta s, \forall s>0.$
	\end{itemize}
\end{remark}
\begin{remark}
	By taking as differential Lyapunov function to $V(\vec{x},\delta\vec{x})=\frac{1}{2}\delta\vec{x}^{\top}\boldsymbol{\Pi}(\vec{x},t)\delta\vec{x}$, with $\boldsymbol{\Pi(\vec{x},t)}$ a smooth Riemannian metric and uniform in $t$, expression \eqref{eq:finslerlyapunovinequality} results in the so-called \emph{generalized contraction analysis}   in \cite{slotinecontraction}, i.e., 
	\begin{equation}
	\frac{\partial \boldsymbol{\Pi}}{\partial \vec{x}}\vec{F}(\vec{x},t)+\frac{\partial \vec{F}^{\top}}{\partial\vec{x}}\boldsymbol{\Pi}(\vec{x},t)+\boldsymbol{\Pi}(\vec{x},t)\frac{\partial \vec{F}}{\partial \vec{x}}< -2\beta\boldsymbol{\Pi}.
	\label{eq:slotinecontractionconditions}
	\end{equation}
\end{remark}

If the interest is  convergence to a specific trajectory, rather than convergence between any two arbitrary trajectories. The concept introduced by \citealp{wang},  with the name of \emph{partial contraction}, gives a solution
\begin{theorem}\label{theo:partialcontraction}
	Assume \eqref{eq:systemcontraction1} is written as the actual system 
	\begin{equation}
	\dot{\vec{x}}=\vec{h}(\vec{x},\vec{x},t),
	\label{eq:actualsystemTheorem}	
	\end{equation}
	Consider a virtual system of the form
	\begin{equation}
	\dot{\vec{x}}_v=\vec{h}(\vec{x}_v,\vec{x},t).
	\label{eq:virtualsystemTheorem}
	\end{equation}
	such that both $\vec{x}_v=\vec{x}$ and a smooth trajectory $\vec{x}_v=\vec{x}_d(t)$ are particular solutions to \eqref{eq:virtualsystemTheorem}. If the virtual system  is contracting, uniformly in $\vec{x}$ and $t$, then $\vec{x}$ (asymptotically/exponentially) converges to $\vec{x}_d(t)$. System \eqref{eq:actualsystemTheorem}  is said to be\emph{ partially contracting}.
\end{theorem}

\begin{remark}\label{remark:Partialcontraction}
	From a control design perspective, we want the control system  \eqref{eq:controlsystemSliding} to track a given desired trajectory $\vec{x}_d$. To that end, also in \cite{wang} and \cite{jouffroy}, an adaptation of Theorem \ref{theo:partialcontraction} was presented as follows. Suppose the control system  \eqref{eq:controlsystemSliding} is rewritten as the actual system 
	\begin{equation}
	\dot{\vec{x}}=\vec{h}(\vec{x},\vec{x},\vec{u},t),
	\label{eq:actualcontrolsystemTheorem}	
	\end{equation}
	and assume that the control input $\vec{u}=\vec{u}(\vec{x},\vec{x}_d,\dot{\vec{x}}_d)$ can be chosen such that
	\begin{equation}
	\dot{\vec{x}}_d=\vec{h}(\vec{x}_d,\vec{x},\vec{u},t),
	\label{eq:desiredcontrolsystemTheorem}	
	\end{equation}	
	where $\vec{x}_d$ is a desired trajectory. Consider now as virtual system to
	\begin{equation}
	\dot{\vec{x}}_v=\vec{h}(\vec{x}_v,\vec{x},\vec{u},t).
	\label{eq:virtualcontrolsystemTheorem}
	\end{equation}
	If system \eqref{eq:virtualcontrolsystemTheorem} is contracting uniformly in $\vec{x}$ and $t$, then conclusion of Theorem \ref{theo:partialcontraction} holds. 
\end{remark}
\subsection{Mechanical port-Hamiltonian systems}
Consider the \emph{input-state port-Hamiltonian} (\cite{pHbook}) representation of a fully-actuated mechanical system of the form 
\begin{equation}
\begin{split}
\begin{bmatrix}
\dot{\vec{q}}\\
\dot{\vec{p}}
\end{bmatrix}&=\begin{bmatrix}
\vec{0}_n & \vec{I}_n\\
-\vec{I}_n& -\vec{D}(\vec{\vec{q}})
\end{bmatrix}\begin{bmatrix}
\frac{\partial H}{\partial \vec{q}}(\vec{q},\vec{p})\\
\frac{\partial H}{\partial \vec{p}}(\vec{q},\vec{p})
\end{bmatrix}+\begin{bmatrix}
\vec{0}_n\\
\vec{G}(\vec{q})
\end{bmatrix}\vec{u},
\end{split}
\label{eq:phmechanical}
\end{equation}
where $\vec{x}=[\vec{q},\vec{p}]^{\top}\in T^*\mathcal{Q}=\mathcal{X}$, and the generalized momentum is $\vec{p}=\vec{M}(\vec{q})\dot{\vec{q}}$, with $\dot{\vec{q}}$ the generalized velocity;  $m_1\vec{I}_n\leq\vec{M}(\vec{q})=\vec{M}^{\top}(\vec{q})\leq m_2\vec{I}_n$ is the bounded inertia matrix, where $m_1,m_2\in\mathds{R}_{>0}$; $\vec{D}(\vec{q})=\vec{D}^{\top}(\vec{q}) \geq \vec{0}_n$ is the damping matrix, the matrices identity  $\vec{I}_n $ and zero $\vec{0}_n$ have dimension $n=\text{dim}\mathcal{Q}$; $\vec{G}(\vec{q})$ is the full-rank input matrix, $\vec{u}\in\mathds{R}^{n}$ the control input and the Hamiltonian function is given by the total energy 
\begin{equation}
H(\vec{x})=\frac{1}{2}\vec{p}^{\top}\vec{M}^{-1}(\vec{q})\vec{p}+V(\vec{q}),
\label{eq:phmechanicalHamiltonian}
\end{equation}
with $V(\vec{q})$ the potential energy.
In \cite{Arimoto} it was proven that the matrix  $\vec{S}(\vec{q},\vec{\dot{q}})$ (which is skew-symmetric, homogeneous and linear in $\dot{\vec{q}}$), defined by
\begin{equation}
{S}_{ij}(\vec{q},\vec{\dot{q}}):=\frac{1}{2}\left[\sum_{k=1}^{n}\dot{q}_k\left(\frac{\partial M_{ik}}{\partial q_j} (\vec{q})-\frac{\partial  M_{jk}}{\partial q_i}(\vec{q})\right)\right],
\end{equation}
where $S_{ij}=-S_{ji}$, fulfills the property
\begin{equation}
\left[\vec{S}(\vec{q},\vec{\dot{q}})-\frac{1}{2}\dot{\vec{M}}(\vec{q})\right]\vec{\dot{q}}=-\frac{\partial}{\partial \vec{q}}\left[\frac{1}{2}\dot{\vec{q}}^{\top}\vec{M}(\vec{q})\dot{\vec{q}}\right].
\label{eq:coriolislagrange}
\end{equation}
Such a property is for the Euler-Lagrange realization of mechanical systems with state variables $(\vec{q},\dot{\vec{q}})$. However, as was shown in \cite{tracking2016}, by applying the \emph{Legendre transformation}, property \eqref{eq:coriolislagrange} can be expressed in state $\vec{x}$ of the Hamiltonian realization \eqref{eq:phmechanical} as
\begin{equation}
\left[\vec{S}(\vec{q},\vec{{p}})-\frac{1}{2}\dot{\vec{M}}(\vec{q})\right]\vec{M}^{-1}(\vec{q})\vec{p}=\frac{\partial }{\partial \vec{q}}\left[\frac{1}{2}{\vec{p}}^{\top}\vec{M}^{-1}(\vec{q}){\vec{p}}\right].
\label{eq:coriolishamilton}
\end{equation}

Using  \eqref{eq:coriolishamilton}, system \eqref{eq:phmechanical}  can be rewritten as
\begin{equation}
\begin{split}
\begin{bmatrix}
\dot{\vec{q}}\\
\dot{\vec{p}}
\end{bmatrix}&=\begin{bmatrix}
\vec{0}_n & \vec{I}_n\\
-\vec{I}_n& -\vec{E}(\vec{q},\vec{p})
\end{bmatrix}\begin{bmatrix}
\frac{\partial V}{\partial \vec{q}}(\vec{q})\hfill \\
\frac{\partial H}{\partial \vec{p}}(\vec{q},\vec{p})
\end{bmatrix}+\begin{bmatrix}
\vec{0}\\
\vec{G}(\vec{q})
\end{bmatrix}\vec{u}
\end{split}
\label{eq:phmechanica3}
\end{equation}
with $\vec{E}(\vec{q},\vec{p}):=\vec{S}(\vec{q},\vec{p})-\frac{1}{2}\dot{\vec{M}}(\vec{q})+\vec{D}(\vec{q})$. Notice that   system \eqref{eq:phmechanica3} preserves passivity, with storage function \eqref{eq:phmechanicalHamiltonian}.
\textcolor{black}{\begin{remark}\label{remark:Coriolis}
	System \eqref{eq:phmechanica3}  can be understood as a port-Hamiltonian system with constant-like inertia matrix. Thus, the constant-like inertia matrix serves as an operator between $T\mathcal{Q}$ and $T^{*}\mathcal{Q}$, i.e., $\vec{p}=\vec{M}^{-1}(\vec{q})\dot{\vec{q}}$, but its dependency on $\vec{q}$ induces an endogeneous disturbance $\boldsymbol{\Phi}(\vec{x})$ in the dissipation matrix\footnote{This due to the fact that at least for the Coulomb friction constant the units are $N\cdot m\cdot s=Kg\cdot m^2/s$  and inertia physical units are $[\vec{M}]=Kg\cdot m^2$. Thus, the derivative $[\dot{\vec{M}}]=Kg\cdot m^2/s$.} without explicit sign definiteness. Explicitly, we have
		\begin{equation}
		\begin{split}
		\dot{\vec{x}}&=\left[\vec{J}(\vec{x})-(\vec{R}(\vec{q})-\Phi(\vec{x}))\right]\begin{bmatrix}
			\frac{\partial V}{\partial \vec{q}}(\vec{q})\hfill \\
			\frac{\partial H}{\partial \vec{p}}(\vec{q},\vec{p})
		\end{bmatrix}+\begin{bmatrix}
		\vec{0}\\
		\vec{G}
		\end{bmatrix}\vec{u}
		\end{split}
		\label{eq:phmechanica4}
		\end{equation}	
		where $\vec{J}=-\vec{J}^{\top}, \vec{R}=\vec{R}^{\top}>\vec{0}_n$  and
		\begin{equation}
		\vec{J}=\begin{bmatrix}
		\vec{0}_n & \vec{I}_n\\
		-\vec{I}_n & -\vec{S}
		\end{bmatrix}, \vec{R}=\begin{bmatrix}
		\vec{0}_n & \vec{0}_n\\
		\vec{0}_m & \vec{D}
		\end{bmatrix}, \boldsymbol{\Phi}=\begin{bmatrix}
		\vec{0}_n & \vec{0}_n\\
		\vec{0}_n & \frac{1}{2}\dot{\vec{M}}
		\end{bmatrix}.
		\end{equation}
		As it was mentioned, system \eqref{eq:phmechanica4} preserves passivity.  Thus,  it is not necessary to compensate by control the disturbance $\Phi(\vec{x})$, since it has not effect on the stability.
	\end{remark}
	}

\section{Trajectory tracking controller}\label{sIII}
\emph{Control objective:}  Design a control law for system \eqref{eq:phmechanical} such that $\vec{x}$ converges to a  smooth desired trajectory $\vec{x}_d(t)$.

To solve the control problem, it is necessary to construct a suitable error system for  \eqref{eq:phmechanical} as in \cite{fujimoto}.
Consider a twice differentiable desired trajectory $\vec{x}_d(t)=[\vec{q}_d(t),\vec{p}_d(t)]^{\top}$, with $\vec{p}_d(t)=\vec{M}(\vec{q}_d(t))\dot{\vec{q}}_d(t)$ and the change of coordinates
	\begin{equation}
	\tilde{\vec{x}}:=\begin{bmatrix}
	\tilde{\vec{q}}\\
	\boldsymbol{\sigma}
	\end{bmatrix}=\begin{bmatrix}
	\vec{q}-\vec{q}_d(t)\\
	\vec{p}-\vec{p}_r(t)
	\end{bmatrix}
	\label{eq:changeofcoordinatesError}
	\end{equation}
	where $\vec{p}_r$ is an auxiliary momentum reference  to be defined. 
The dynamics of $\tilde{\vec{q}}$ in \eqref{eq:changeofcoordinatesError} is 
\begin{equation}
\dot{\tilde{\vec{q}}}=\vec{M}^{-1}(\tilde{\vec{q}}+\vec{q}_d)\vec{p}-\vec{M}^{-1}(\vec{q}_d)\vec{p}_d.
\label{eq:dottildaq1}
\end{equation}
due to space limitations, we define the following notation $\vec{M}(\tilde{\vec{q}}+\vec{q}_d)=\vec{M}(\tilde{\vec{q}},t)$.

Like in \emph{backstepping}, assume $\vec{p}=\boldsymbol{\sigma}+\vec{p}_r$ is a control input to \eqref{eq:dottildaq1}, with $\boldsymbol{\sigma}$ as new state and $\vec{p}_r$ as a stabilizing term. After substitution of $\vec{p}$ and defining to $\vec{p}_r$ as
	\begin{equation}
	\vec{p}_r=\vec{p}_{d\sigma}-\boldsymbol{\Lambda}\tilde{\vec{q}},
	\label{eq:auxiliarreference}
	\end{equation}
	for $\vec{p}_{d\sigma}=\vec{M}(\tilde{\vec{q}},t)\dot{\vec{q}}_d$, $-\boldsymbol{\Lambda}=-\boldsymbol{\Lambda}^{\top}$ is a Hurwitz matrix. It results in the position error dynamics
\begin{equation}
\dot{\tilde{\vec{q}}}=\vec{M}^{-1}(\tilde{\vec{q}},t)\left(\boldsymbol{\sigma}-\boldsymbol{\Lambda}\tilde{\vec{q}}\right),
\label{eq:closed-loop position error system}
\end{equation}
with $\boldsymbol{\sigma}$ as input. When $\boldsymbol{\sigma}=\vec{0}$ in \eqref{eq:closed-loop position error system},  the origin $\tilde{\vec{q}}=\vec{0}$ is asymptoticly stable, since $-\vec{M}^{-1}(\tilde{\vec{q}},t)\boldsymbol{\Lambda}$ is a Hurwitz matrix\footnote{It was proven in Theorem 3.1 of \cite{hurwitz}. Using the paper's notation, take $G_s=-\vec{M}^{-1}(\vec{q})$ and $L=\boldsymbol{\Lambda}$.}. Above implies ${\vec{q}}\rightarrow \vec{q}_d$ as $t\rightarrow \infty$. Simultaneously, from \eqref{eq:auxiliarreference}, $\vec{p}_r\rightarrow\vec{p}_d$ as $t\rightarrow \infty$.

The dynamics of $\boldsymbol{\sigma}$ is $\dot{\boldsymbol{\sigma}}=\dot{\vec{p}}-\dot{\vec{p}}_r$ evaluated in the change of coordinates \eqref{eq:changeofcoordinatesError}.
 Then, an error system for \eqref{eq:phmechanical}  is
 \begin{equation}
 \begin{split}
 \dot{\tilde{\vec{q}}}&=\vec{M}^{-1}(\tilde{\vec{q}},t)\left(\boldsymbol{\sigma}-\boldsymbol{\Lambda}\tilde{\vec{q}}\right)\\
 \dot{\boldsymbol{\sigma}}&=-\left[\frac{\partial H}{\partial \vec{q}}(\vec{x})+\vec{D}(\vec{q})\frac{\partial H}{\partial \vec{p}}(\vec{x})-\vec{G}(\vec{q})\vec{u}+\dot{\vec{p}}_r\right]_{\tiny
 	\begin{array}{c l}
 	\vec{q}=\tilde{\vec{q}}+\vec{q}_d\\
 	\vec{p}=\boldsymbol{\sigma}+\vec{p}_r
 	\end{array}}
 \end{split}
 \label{eq:errorsystemPHS}
 \end{equation}


\vspace{-0.3cm}
The following result gives a solution to the control problem. For sake of space, some arguments are left out.
\begin{proposition}
	Consider a twice differentiable desired  trajectory $\vec{x}_d\in T^*\mathcal{Q}$, together with the change of coordinates \eqref{eq:changeofcoordinatesError} and \eqref{eq:auxiliarreference}. Consider also the  pH system \eqref{eq:phmechanical} in closed-loop with the control law\footnote{With $\frac{\partial}{\partial \vec{q}}\left(\vec{p}_r^{\top}\vec{M}^{-1}(\vec{q})\boldsymbol{\sigma}\right)=\vec{S}(\vec{q},\boldsymbol{\sigma})\frac{\partial H}{\partial \vec{p}}(\vec{q},\vec{p}_r)+\vec{S}(\vec{q},\vec{p}_r)\frac{\partial H}{\partial \vec{p}}({\vec{q}},\boldsymbol{\sigma})$.}
	\begin{equation}
	\begin{split}
	\vec{G}\vec{u}&=\vec{G}\vec{u}_{eq}+\vec{G}\vec{u}_{at},\\	
	\vec{G}\vec{u}_{eq}&=\dot{\vec{p}}_r+\frac{\partial H}{\partial \vec{q}}(\vec{q},\vec{p}_r)+\vec{D}(\vec{q})\frac{\partial H}{\partial \vec{p}}(\vec{q},\vec{p}_r),\\
	\vec{G}\vec{u}_{at}&=-\vec{K}_d\frac{\partial H}{\partial \vec{p}}({\vec{q}},\boldsymbol{\sigma})-\vec{M}^{-1}(\vec{q})\boldsymbol{\Lambda}\tilde{\vec{q}}+\frac{\partial}{\partial \vec{q}}\left(\vec{p}_r^{\top}\vec{M}^{-1}(\vec{q})\boldsymbol{\sigma}\right),
	\end{split}
	\label{eq:controlaw}
	\end{equation}
	where 
	$\vec{K}_d$ fulfills
	\begin{equation}
	\vec{D}+\vec{K}_d+\frac{1}{2}\vec{I}_n-\frac{1}{4}(\vec{M}^{-1}+\vec{M})>0.
	\label{eq:contractionShuur}
	\end{equation}

	 Then,
	\begin{enumerate}
		\item The closed-loop system in error coordinates \eqref{eq:changeofcoordinatesError} is given by
	\begin{equation}
	\dot{\tilde{\vec{x}}}=\begin{bmatrix}
	-\vec{I} & \vec{I}\\
	-\vec{I} & -\vec{E}(\tilde{\vec{q}}+\vec{q}_d,\boldsymbol{\sigma})-\vec{K}_d
	\end{bmatrix}		\begin{bmatrix}
	\vec{M}^{-1}(\tilde{\vec{q}},t)\boldsymbol{\Lambda}\tilde{\vec{q}}\\
	\vec{M}^{-1}(\tilde{\vec{q}},t)\boldsymbol{\sigma}
	\end{bmatrix}.
	\label{eq:closederrorsystemnoconstantMcorrect}
	\end{equation}				


\item The origin of \eqref{eq:closederrorsystemnoconstantMcorrect}  is exponentially stable with rate 
		\begin{equation}
		\beta=\min\text{eig}\left(\vec{P}^{1/2}(\tilde{\vec{x}})\boldsymbol{\varUpsilon}(\tilde{\vec{x}})\vec{P}^{1/2}(\tilde{\vec{x}})\right).
		\label{eq:convergencerate}
		\end{equation} 
	    where $\min\text{eig}(\cdot)$ denotes the minimum eigenvalue of the matrix in the argument and matrices
	\begin{equation}
	\vec{P}(\tilde{\vec{x}})=\begin{bmatrix}
	{\boldsymbol{\Lambda}} & \vec{0}\\
	\vec{0} & {\vec{M}}^{-1}(\tilde{\vec{q}}+\vec{q}_d)
	\end{bmatrix},
	\label{eq:contractionmetricerror}
	\end{equation}
	\begin{equation}
\boldsymbol{\varUpsilon}(\tilde{\vec{x}})=\begin{bmatrix}
2\vec{M}^{-1} & (\vec{M}^{-1}-\vec{I}_n)\\
( \vec{M}^{-1}-\vec{I}_n) & 2(\vec{D}+\vec{K}_d)
\end{bmatrix}.
	\label{eq:Shur}
	\end{equation}	
		\item The sliding manifold 
		\begin{equation}
		\Omega(t)=\{\vec{x}\in T^*\mathcal{Q}: \boldsymbol{\sigma}(\vec{x},t)=\tilde{\vec{p}}_{\sigma}+\boldsymbol{\Lambda}\tilde{\vec{q}}=\vec{0}\},
		\label{eq:slidingmanifold1}
		\end{equation}	
			where $\tilde{\vec{p}}_{\sigma}:=\vec{M}(\vec{q})\dot{\tilde{\vec{q}}}$, is invariant and attractive, for system \eqref{eq:closederrorsystemnoconstantMcorrect}, with ideal sliding motion 
			\begin{equation}
			\dot{\vec{q}}=\frac{\partial H}{\partial \vec{p}}(\vec{q},\vec{p}_r).
			\end{equation}
	\end{enumerate}
\end{proposition}
\emph{Proof:} 
\begin{enumerate}
	\item \textcolor{black}{
	\indent Straight forward computations after substitution of \eqref{eq:controlaw} in the error system \eqref{eq:errorsystemPHS} gives the closed-loop system \eqref{eq:closederrorsystemnoconstantMcorrect}}
	\item To prove this item, we will use partial contraction\footnote{By defining  $\vec{e}=\tilde{\vec{x}}_a$ and $\vec{x}=\tilde{\vec{x}}$,  properties labeled as TULES-NL, UES-TL and ULMTE in \cite{andrieu2016}, are also verified.} Theorem \ref{theo:partialcontraction}. Consider the following virtual system with state $\tilde{\vec{x}}_a=[\tilde{\vec{q}}_a^{\top},\boldsymbol{\sigma}_a^{\top}]^{\top}$
	\begin{equation}
	\dot{\tilde{\vec{x}}}_a=\begin{bmatrix}
	-\vec{I} & \vec{I}\\
	-\vec{I} & -(\vec{E}(t,\boldsymbol{\sigma})+\vec{K}_d)
	\end{bmatrix}\begin{bmatrix}
	\vec{M}^{-1}\boldsymbol{\Lambda\tilde{\vec{q}}}_a\\
	\vec{M}^{-1}\boldsymbol{\sigma}_a\\
	\end{bmatrix}.
	\label{eq:auxclosederrorsystemnoconstantM}
	\end{equation}
	Notice $\tilde{\vec{x}}_a=\tilde{\vec{x}}$  and $\tilde{\vec{x}}_a=\vec{0}$ are two  particular solutions of system \eqref{eq:auxclosederrorsystemnoconstantM}. The variational dynamics of the virtual system \eqref{eq:auxclosederrorsystemnoconstantM} is 
	\begin{equation}
	\begin{split}
	\delta\dot{\tilde{\vec{x}}}_a&=-\begin{bmatrix}
	\vec{M}^{-1}\boldsymbol{\Lambda} & -\vec{M}^{-1}\\
	\vec{M}^{-1}\boldsymbol{\Lambda} & (\vec{E}+\vec{K}_d)\vec{M}^{-1}
	\end{bmatrix}\delta\tilde{\vec{x}}_a.	
	\end{split}
	\label{eq:variationalauxclosederrorsystemnoconstantM}
	\end{equation}	
	For the prolonged system \eqref{eq:auxclosederrorsystemnoconstantM}-\eqref{eq:variationalauxclosederrorsystemnoconstantM}, let the candidate differential Lyapunov function be
	\begin{equation}
	V(\tilde{\vec{x}}_a,\delta\tilde{\vec{x}}_a,t)=\frac{1}{2}	\delta\tilde{\vec{x}}_a^{\top}\vec{P}(\tilde{\vec{x}})	\delta\tilde{\vec{x}}_a.
	\label{eq:dlyapunov}
	\end{equation}
	The time derivative of \eqref{eq:dlyapunov} is 	\vspace{-0.1cm}
	\begin{equation}
	\begin{split}
	\dot{V}&=-\delta\tilde{\vec{x}}_a^{\top}\begin{bmatrix}
	\boldsymbol{\Lambda}\vec{M}^{-1}\boldsymbol{\Lambda} & -\boldsymbol{\Lambda}\vec{M}^{-1}\\
	\vec{M}^{-2}\boldsymbol{\Lambda} & \vec{M}^{-1}(\vec{E}+\vec{K}_d)\vec{M}^{-1}
	\end{bmatrix}\delta\tilde{\vec{x}}_a\\&\hspace{1cm}+\frac{1}{2}\delta\tilde{\vec{x}}_a^{\top}\begin{bmatrix}
	\dot{\boldsymbol{\Lambda}} & \vec{0}\\
	\vec{0} & \dot{\vec{M}}^{-1}(\vec{q})
	\end{bmatrix}\delta\tilde{\vec{x}}_a,\\
	&=-\delta\tilde{\vec{x}}_a^{\top}\begin{bmatrix}
	\boldsymbol{\Lambda}\vec{M}^{-1}\boldsymbol{\Lambda} & -\boldsymbol{\Lambda}\vec{M}^{-1}\\
	\vec{M}^{-2}\boldsymbol{\Lambda} & \vec{M}^{-1}(\vec{D}+\vec{K}_d)\vec{M}^{-1}
	\end{bmatrix}\delta\tilde{\vec{x}}_a\\&\hspace{0.5cm}+
	\delta\boldsymbol{\sigma}_a
	^{\top}\left[ -\vec{M}^{-1}(\vec{S}-\frac{1}{2}\dot{\vec{M}})\vec{M}^{-1}+\frac{1}{2}\dot{\vec{M}}^{-1}
	\right]	\delta\boldsymbol{\sigma}_a,\\
	&=-\delta\tilde{\vec{x}}_a^{\top}\underbrace{\begin{bmatrix}
	\boldsymbol{\Lambda}\vec{M}^{-1}\boldsymbol{\Lambda} & -\boldsymbol{\Lambda}\vec{M}^{-1}\\
	\vec{M}^{-2}\boldsymbol{\Lambda} & \vec{M}^{-1}(\vec{D}+\vec{K}_d)\vec{M}^{-1}
	\end{bmatrix}}_{\boldsymbol{\Xi}(\tilde{\vec{x}})}\delta\tilde{\vec{x}}_a,\\ 
	\end{split}
	\label{eq:ddiffLyapunov}
	\end{equation}		
where the symmetric part of $\boldsymbol{\Xi}(\tilde{\vec{x}})$ is expressed as 
\begin{equation}
\text{Sym}(\boldsymbol{\Xi}(\tilde{\vec{x}}))=\frac{1}{2}
\vec{P}(\tilde{\vec{x}})
\boldsymbol{\varUpsilon}(\tilde{\vec{x}})
\vec{P}(\tilde{\vec{x}})
\label{eq:ddiffLyapunovSchurr}
\end{equation}
Thus, \eqref{eq:ddiffLyapunov} will be negative definite if and only if  the Schur complement of matrix \eqref{eq:Shur} with respect to $2\vec{M}^{-1}$ fulfills  \eqref{eq:contractionShuur}. Which is always possible by choosing a big enough $\vec{K}_d$. 
Therefore, the prolonged system \eqref{eq:auxclosederrorsystemnoconstantM}-\eqref{eq:variationalauxclosederrorsystemnoconstantM} contracts  \eqref{eq:dlyapunov} with respect to the metric \eqref{eq:contractionmetricerror} in $T\mathcal{X}$. 

With \eqref{eq:convergencerate}, the time derivative \eqref{eq:ddiffLyapunov} satisfies 
\begin{equation}
\dot{V}(\tilde{\vec{x}}_a,\delta \tilde{\vec{x}}_a,t)<- 2\beta V(\tilde{\vec{x}}_a,\delta \tilde{\vec{x}}_a,t)
\end{equation}
uniformly in $t$ and $\tilde{\vec{x}}$, or equivalently \eqref{eq:slotinecontractionconditions} for the matrix \eqref{eq:contractionmetricerror}. By Remark \ref{remark:incremental}, the virtual system \eqref{eq:auxclosederrorsystemnoconstantM} is incrementally exponentially stable with rate \eqref{eq:convergencerate}. Therefore,  $\tilde{\vec{x}}$ converges to $\vec{0}$ exponentially with rate $\beta$, as $t\rightarrow \infty$.

\item The existence of $\vec{u}_{eq}$ in \eqref{eq:controlaw}, guarantees that the sliding manifold \eqref{eq:slidingmanifold1} is rendered to an invariant manifold. From the previous item,  $\boldsymbol{\sigma}\rightarrow \vec{0}$, which means the invariant manifold is attractive.
Finally, system \eqref{eq:phmechanical} has \emph{regular canonical form}, and definitions of $\vec{p}_r$ and $\boldsymbol{\sigma}$  imply, by straightforward computations, the \emph{reduced-order} ideal sliding motion
\begin{equation}
\dot{\vec{q}}=\frac{\partial H}{\partial \vec{p}}(\vec{q},\vec{p}_r).
\end{equation}
\hfill$\blacksquare$
\end{enumerate}

In \cite{sanfeliceconvergence2}, an observer was designed for shrinking a Riemannian distance, instead of designing a contracting observer. The following proposition shows that the controller \eqref{eq:controlaw} has the same property.
\begin{proposition}
	Consider system \eqref{eq:phmechanical}. The control law \eqref{eq:controlaw} shrinks the Riemannian distance $d(\vec{x},\vec{x}_d)$ induced by
	\begin{equation}
	\boldsymbol{\Pi}(\vec{x})=\begin{bmatrix}
	\boldsymbol{\Lambda}+\boldsymbol{\Lambda}\vec{M}^{-1}(\vec{q})\boldsymbol{\Lambda} & \boldsymbol{\Lambda}\vec{M}^{-1}(\vec{q})\\
	\vec{M}^{-1}(\vec{q})\boldsymbol{\Lambda} & \vec{M}^{-1}(\vec{q})
	\end{bmatrix}.
	\label{eq:contractionmetricPHS}
	\end{equation}
\end{proposition}
\textcolor{black}{
	\emph{Proof:} We will show that both, the actual system \eqref{eq:phmechanical} and the system driven by controller \eqref{eq:controlaw}, are partially contracting to a virtual system with respect to the metric \eqref{eq:contractionmetricPHS}, in the sense of Remark \ref{remark:Partialcontraction}. This means, that in particular, the distance induced by the metric \eqref{remark:Partialcontraction}, between the actual state  $\vec{x}$ and the desired state $\vec{x}_d$ shrinks.\\	
	To that end, fist we express the controller \eqref{eq:controlaw} in implicit form for the state $\vec{x}_{d\sigma}=[\vec{q}_d,\vec{p}_{d\sigma}]^{\top}$ in original coordinates $\vec{x}$. Using \eqref{eq:auxiliarreference} and $\boldsymbol{\sigma}=\tilde{\vec{p}}_{\boldsymbol{\sigma}}+\boldsymbol{\Lambda}\tilde{\vec{a}}$, we have
	\begin{equation}
	\begin{split}
	&\vec{G}(\vec{q})\vec{u}=\frac{\partial V}{\partial \vec{q}}(\vec{q})+\left[\vec{E}(\vec{q},\vec{p}_{d\boldsymbol{\sigma}})+\vec{S}(\vec{q},\tilde{\vec{p}}_{\boldsymbol{\sigma}})\right]\vec{M}^{-1}(\vec{q})\vec{p}_{d\boldsymbol{\sigma}}\\&-\vec{A}(\tilde{\vec{q}},\tilde{\vec{p}}_{\sigma})\vec{M}^{-1}(\vec{q})\boldsymbol{\Lambda}\tilde{\vec{q}}-\vec{B}(\tilde{\vec{q}},\tilde{\vec{p}}_{\sigma})\vec{M}^{-1}(\vec{q})\tilde{\vec{p}}_{\sigma}+\dot{\vec{p}}_{d\boldsymbol{\sigma}}.
	\end{split}
	\end{equation}
	where 
	\begin{equation}
	\begin{split}
	\vec{A}(\tilde{\vec{q}},\tilde{\vec{p}}_{\sigma})&:=\vec{E}(\vec{q},\tilde{\vec{p}}_{\boldsymbol{\sigma}}+\boldsymbol{\Lambda}\tilde{\vec{q}})+\vec{K}_d+\vec{I},\\
	\vec{B}(\tilde{\vec{q}},\tilde{\vec{p}}_{\sigma})&:=\boldsymbol{\Lambda}+\vec{K}_d-\vec{S}(\vec{q},\vec{p}_{d\boldsymbol{\sigma}}-\boldsymbol{\Lambda}\tilde{\vec{q}}).
	\end{split}
	\end{equation}	
	Considering the fact that $\dot{\vec{q}}_d=\vec{M}^{-1}(\vec{q})\vec{p}_{d\boldsymbol{\sigma}}$, we can rewrite the controller as
	\begin{equation}
	\begin{split}
	\dot{\vec{x}}_{d\sigma}&=\begin{bmatrix}
	\vec{0} & \vec{I}\\
	-\vec{I} & -\vec{E}(\vec{q},\vec{p}_{d\boldsymbol{\sigma}}+\tilde{\vec{p}}_{\sigma})
	\end{bmatrix}\begin{bmatrix}
	\frac{\partial V}{\partial \vec{q}}(\vec{q})\hfill\\
	\vec{M}^{-1}(\vec{q}){\vec{p}}_{d\sigma}
	\end{bmatrix}+\begin{bmatrix}
	\vec{0}\\
	\vec{G}
	\end{bmatrix}\vec{u}\\&+\begin{bmatrix}
	\vec{0} & \vec{0}\\
	\vec{A}(\tilde{\vec{q}},\tilde{\vec{p}}_{\sigma}) &\vec{B}(\tilde{\vec{q}},\tilde{\vec{p}}_{\sigma})
	\end{bmatrix}\begin{bmatrix} 
	\vec{M}^{-1}(\vec{q})\boldsymbol{\Lambda}\tilde{\vec{q}}  \\
	\vec{M}^{-1}(\vec{q})\tilde{\vec{p}}_{\sigma}
	\end{bmatrix},\\
	&=\begin{bmatrix}
	\vec{0} & \vec{I}\\
	-\vec{I} & -\vec{D}-\vec{S}(\vec{q},\tilde{\vec{p}}_{\sigma})
	\end{bmatrix}	\frac{\partial H}{\partial \vec{x}}(\vec{q},\vec{p}_{d\sigma})+\begin{bmatrix}
	\vec{0}\\
	\vec{G}
	\end{bmatrix}\vec{u}\\&+\begin{bmatrix}
	\vec{0} & \vec{0}\\
	\vec{A}(\tilde{\vec{q}},\tilde{\vec{p}}_{\sigma}) &\vec{B}(\tilde{\vec{q}},\tilde{\vec{p}}_{\sigma})
	\end{bmatrix}\begin{bmatrix} 
	\vec{M}^{-1}(\vec{q})\boldsymbol{\Lambda}\tilde{\vec{q}}  \\
	\vec{M}^{-1}(\vec{q})\tilde{\vec{p}}_{\sigma}
	\end{bmatrix}.
	\end{split}
	\label{eq:controllerNonconstantM}
	\end{equation} 
	\newline
	Let a virtual system with state $\vec{x}_v=[\vec{q}_v,\vec{p}_v]^{\top}$ be
	\begin{equation}
	\begin{split}
	\dot{\vec{x}}_{v}&=\begin{bmatrix}
	\vec{0} & \vec{I}\\
	-\vec{I} & -\vec{E}(\vec{q},\vec{p}_{d\sigma}+\tilde{\vec{p}}_{\sigma})
	\end{bmatrix}	\begin{bmatrix}
	\frac{\partial V}{\partial \vec{q}}(\vec{q})\hfill\\
	\vec{M}^{-1}(\vec{q}){\vec{p}}_{v}
	\end{bmatrix}+\begin{bmatrix}
	\vec{0}\\
	\vec{G}
	\end{bmatrix}\vec{u}\\&+\begin{bmatrix}
	\vec{0} & \vec{0}\\
	\vec{A}(\tilde{\vec{q}},\tilde{\vec{p}}_{\sigma}) &\vec{B}(\tilde{\vec{q}},\tilde{\vec{p}}_{\sigma})
	\end{bmatrix}\begin{bmatrix} 
	\vec{M}^{-1}(\vec{q})\boldsymbol{\Lambda}\tilde{\vec{q}}_v  \\
	\vec{M}^{-1}(\vec{q})\tilde{\vec{p}}_{v}
	\end{bmatrix},
	\end{split}
	\label{eq:virtualNonconstantM}
	\end{equation} 
	with $\tilde{\vec{q}}_v=\vec{q}-\vec{q}_v$, $\tilde{\vec{p}}_v=\vec{p}-\vec{p}_v$ and  $\tilde{\vec{x}}^{\top}_v=[\tilde{\vec{q}}_v.\tilde{\vec{p}}_v]$. Notice  that system \eqref{eq:virtualNonconstantM} has as particular solutions to both,  $\vec{x}_v=\vec{x}_{d\sigma}$ and $\vec{x}_v=\vec{x}$. 	The variational system of \eqref{eq:virtualNonconstantM} is
	\begin{equation}
	\begin{split}
	\delta\dot{\vec{x}}_v&=-\begin{bmatrix}
	\vec{0} & -\vec{M}^{-1}\\
	\vec{A}\vec{M}^{-1}\boldsymbol{\Lambda} &\left(\vec{E}+\vec{B}\right)\vec{M}^{-1}
	\end{bmatrix}\delta\vec{x}_v.
	\end{split}
	\label{eq:variationalvirtualNonconstantM}
	\end{equation}}

Now, consider the following change of coordinates
\begin{equation}
\delta\tilde{\vec{x}}_a=-\begin{bmatrix}
\vec{I} & \vec{0}\\
\boldsymbol{\Lambda} &\vec{I}
\end{bmatrix}\delta\vec{x}_v=-\boldsymbol{\Theta}\delta\vec{x}_v.
\label{eq:changeofcoordinates}
\end{equation}
Then, system \eqref{eq:variationalvirtualNonconstantM} in the new coordinates is 
\begin{equation}
\begin{split}
\delta\dot{\tilde{\vec{x}}}_a=-\boldsymbol{\Theta}\begin{bmatrix}
\vec{0} & -\vec{M}^{-1}\\
\vec{A}\vec{M}^{-1}\boldsymbol{\Lambda} &\left(\vec{E}+\vec{B}\right)\vec{M}^{-1}
\end{bmatrix}\boldsymbol{\Theta}^{-1}\delta\tilde{\vec{x}}_a
\end{split}
\end{equation}
which is nothing but \eqref{eq:variationalauxclosederrorsystemnoconstantM}. Then, the virtual system \eqref{eq:virtualNonconstantM} is contracting with rate $2\beta$, with respect to the differential Lyapunov function or contraction measure 
\begin{equation}
\overline{V}(\vec{x}_v,\delta\vec{x}_v)=\frac{1}{2}\delta\vec{x}_v^{\top}\boldsymbol{\Theta}^{\top}\vec{P}(\tilde{\vec{x}})\boldsymbol{\Theta}\delta\vec{x}_v
\label{eq:difflyapunovvirtualnormal}
\end{equation}
where  $\boldsymbol{\Pi}(\vec{x})=\boldsymbol{\Theta}^{\top}\vec{P}(\tilde{\vec{x}})\boldsymbol{\Theta}$. Thus, $\overline{V}(\vec{x}_v,\delta\vec{x}_v)<e^{-2\beta t}$.

Now, as in \cite{forni}, consider the set $\Gamma(\vec{x},\vec{x}_{d\sigma})$ of all normalized paths $\boldsymbol{\gamma}:[0,1]\rightarrow \mathcal{X}$ connecting $\vec{x}$ with $\vec{x}_d$ such that $\boldsymbol{\gamma}(0)=\vec{x}$ and $\boldsymbol{\gamma}(1)=\vec{x}_d$. Function \eqref{eq:difflyapunovvirtualnormal} defines a \emph{Finsler} structure in $ T\mathcal{X}$, which by integration induces the distance
\begin{equation}
d(\vec{x},\vec{x}_d)=\inf_{\Gamma(\vec{x},\vec{x}_d)}\int_{\gamma}\sqrt{\overline{V}(\boldsymbol{\gamma}(s),\frac{\partial \boldsymbol{\gamma}}{\partial s}(s))}ds<e^{-\beta t}.
\end{equation}
Therefore,  $d(\vec{x},\vec{x}_d)\rightarrow 0$ as $t\rightarrow \infty$ with rate $\beta$.
\hfill $\blacksquare$

\section{Case of study: 3 dof  scara robot}\label{sIV}
Consider a SCARA robot with configuration manifold $\mathcal{Q}=S^1\times S^1 \times \mathds{R}$, and $S^1$ the unitary circumference.
The generalized position vector  $\vec{q}^{\top}=[\theta_1,\theta_2,z]$, generalized momentum $\vec{p}^{\top}=[p_{\theta_1}, p_{\theta_2},p_z]$ and generalized force $\vec{u}^{\top}=[\tau_1, \tau_2,f]$. Such system has a pH representation of form \eqref{eq:phmechanical}, where the inertia matrix is given by  
\begin{equation}
\vec{M}(\vec{q})=\begin{bmatrix}
M_{11}& M_{12}& 0\\
M_{12} &   m_3l_2^2 &               0\\
0  &       0   &  (m_1+m_2+m_3)g
\end{bmatrix},	
\end{equation}
and
\begin{equation*}
\begin{split}
M_{11}&=(m_2+m_3)l_1^2+m_3l_2^2+2m_3l_1l_2\cos\theta_2,\\
M_{12}&= m_3l_2^2+m_3l_1l_2\cos\theta_2.
\end{split}
\end{equation*}
The potential energy is $V(\vec{q})=(m_1+m_2+m_3)gz$, the dissipation  and input matrices are  $\vec{D}=0.2\vec{I}$ and $\vec{G}(\vec{q})=\vec{I}_3$, respectively.

The goal is to track  to $\vec{q}_d=[\sin(t)+1,\sin(t), \sin(t)]^{\top}$,
by closing the loop with the control scheme \eqref{eq:controlaw}, with gain matrices  $\boldsymbol{\Lambda}=\text{diag}\{15,15,15\}$ and $\vec{K}_d=\text{diag}\{30,60,90\}$.

\begin{figure}[h!]
	\vspace{-0.23cm}	  
	\centering
	\includegraphics[width=0.50 \textwidth]{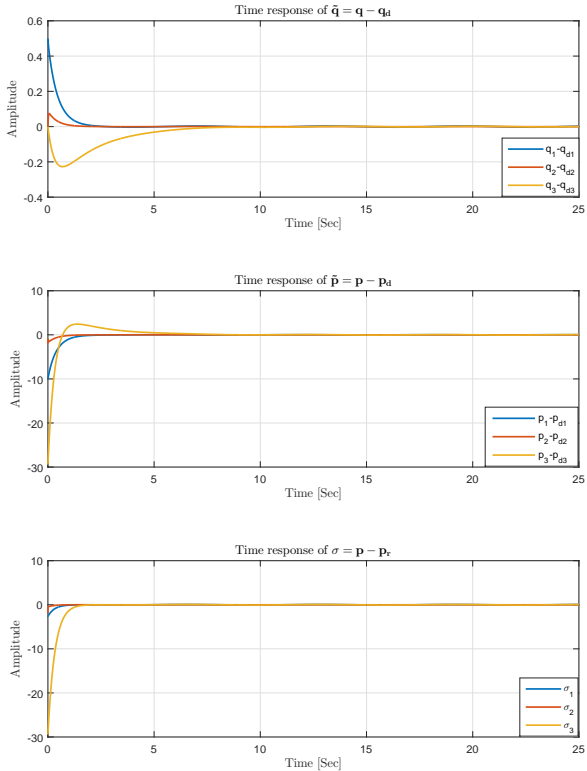}
	\vspace{-1.5cm}
	\caption{Position and momentum error and sliding variable}
	\label{fig:positionerror}	
\end{figure}

In Figure \ref{fig:positionerror}, the time responses of the error variables are shown. All converge to zero exponentially after transients too.  Notice the zero steady-state value of time response of $\tilde{\vec{q}}$  is guaranteed by $\boldsymbol{\sigma}=\vec{0}$. Above is actually the reason due to  $\tilde{\vec{q}}$ and $\tilde{\vec{p}}$ converge slower than the sliding variable.

\begin{figure}[h!]
	\vspace{-0.023cm}		
	\centering
	\includegraphics[width=0.52 \textwidth]{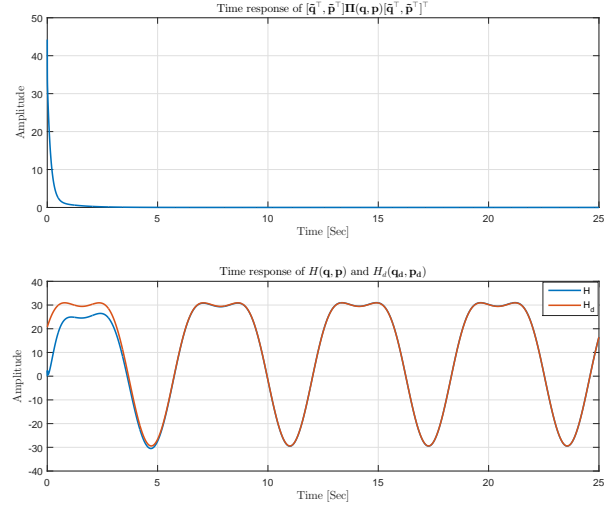}
	\vspace{-1cm}	
	\caption{Contraction measure and Hamiltonian functions}	
	\label{fig:norma}	
\end{figure}

Upper plot of Figure \ref{fig:norma} shows the time response of the contraction measure with respect to the desired trajectory 
(assuming $\boldsymbol{\gamma}$ is a straight line), which after an overshoot transient, in fact shrinks. This is reflected in the lower plot shows the time response of the Hamiltonian versus desired Hamiltonian function. 

\begin{figure}[h!]
	\includegraphics[width=0.52 \textwidth]{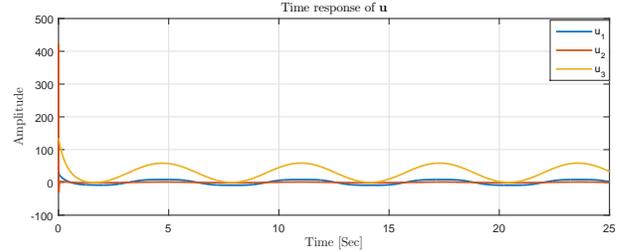}
	\caption{Control signal time response}	
	\label{fig:control}	
\end{figure}

The control effort is shown in Figure \ref{fig:control}. It converges to a steady state trajectory smoothly,  after a big transient. This due to the  \eqref{eq:coriolishamilton} has a big transient, and the controller was required to  compensate it.

\section{Conclusions}\label{sV}

In this paper we presented a trajectory tracking controller  for fully-actuated port-Hamiltonian systems. The control law is composed by the equivalent control, which renders the sliding surface to an invariant set; and a feedback controller which ensured attractivity by making a partially contracting closed-loop error system. Moreover, the controller contracts exponentially a Riemannian distance as result of incremental stability properties of the virtual system. Simulations showed the  good  performance of the proposed controller. 


\section{Acknowledgment}
The fist author is grateful with prof. Romeo Ortega, Alejandro Donaire and Pablo Borja for their constructive comments, which helped to improve the paper presentation.

Also, the first author thanks to the Mexican Consul of Science and Technology and the Government of the State of Puebla for the grant assigned to CVU  $386575$.


\bibliography{bibliografia}                                                     








\appendix
\end{document}